\documentclass[aps,prl,twocolumn,superscriptaddress,nofootinbib]{revtex4-1}

\usepackage{amsmath}
\usepackage{amsfonts}
\usepackage{amssymb}
\usepackage{bm}
\usepackage{xparse}
\usepackage{graphicx}
\usepackage{xcolor}
\usepackage{url}
\usepackage[normalem]{ulem}

\newcommand{\avg}[1]{\langle{#1}\rangle} 
\newcommand{\ket}[1]{| {#1} \rangle} 
\newcommand{\bra}[1]{\langle {#1} |} 
\newcommand{\braket}[2]{\langle {#1} \vphantom{#2} | {#2} \vphantom{#1} \rangle} 
\newcommand{\ketbra}[2]{| {#1} \vphantom{#2} \rangle\langle {#2} \vphantom{#1} |} 
\DeclareDocumentCommand{\Tr}{m m O{\big}}{{\rm Tr}_{\:\!{#1}}#3({#2}#3)}
\DeclareMathOperator*{\ox}{\otimes}

\begin{document}
\title{Macroscopic Superpositions as Quantum Ground States}
\author{Borivoje Daki\'c}
\affiliation{
Institute for Quantum Optics and Quantum Information (IQOQI),
Austrian Academy of Sciences, Boltzmanngasse 3,
A-1090 Vienna, Austria
}
\affiliation{
Faculty of Physics,
University of Vienna,
Boltzmanngasse 5,
A-1090 Vienna, Austria
}
\author{Milan Radonji\'{c}}
\affiliation{
Faculty of Physics,
University of Vienna,
Boltzmanngasse 5,
A-1090 Vienna, Austria
}
\affiliation{
Institute of Physics Belgrade,
University of Belgrade,
Pregrevica 118,
11080 Belgrade, Serbia
}
\date{\today}

\begin{abstract}
We study the question of what kind of a macroscopic superposition can(not) naturally exist as a ground state of some gapped local many-body Hamiltonian. We derive an upper bound on the energy gap of an arbitrary physical Hamiltonian provided that its ground state is a superposition of two well-distinguishable macroscopic ``semiclassical'' states. For a large class of macroscopic superposition states we show that the gap vanishes in the macroscopic limit. This in turn shows that preparation of such states by simple cooling to the ground state is not experimentally feasible and requires a different strategy. Our approach is very general and can be used to rule out a variety of quantum states, some of which do not even exhibit macroscopic quantum properties. Moreover, our methods and results can be used for addressing quantum marginal related problems.
\end{abstract}

\maketitle

\emph{Introduction.}---Ever since Schr\"odinger's cat gedanken experiment \cite{Shroedinger35} the question of whether a
macroscopic system can be found in a quantum superposition state remains unanswered. Various attempts were made to address our
inability to detect macroscopic quantum superpositions. Decoherence-type arguments are commonly employed in which one advocates that the
quantumness of a macroscopic system is lost due to interactions with a noisy environment \cite{Zurek03}. Alternatively, it was
indicated that classical behavior can emerge because our measurements suffer from limited resolution or limited sensitivity
\cite{Brukner07,Simon11,Gisin14}. Moreover, various spontaneous collapse models introduce a stochastic nonlinear modification
of the Schr\"odinger equation that causes macroscopic superpositions to quickly appear as classical, while giving the same
experimental predictions as quantum theory in the microscopic regime \cite{Bassi13}.

Naturally, the boundary between the quantum and classical realms should be explored by experiments \cite{Leggett02,Arndt14,Vedral15}.
In recent decades, typical quantum features have been demonstrated in large molecules \cite{Zeilinger03,Eibenberger13}, hundreds
of photons \cite{Bruno13,Lvovsky13}, superconducting circuits \cite{VanDerWal00,Friedman00}, micromechanical oscillators
\cite{Connell10,Kiesel13}, and fragmented Bose condensates \cite{Fischer09,Fischer15}. Nonetheless, quantum superpositions of truly macroscopic objects remain an uncharted territory that will hopefully be revealed by future experiments.

Recently, different measures have been proposed to quantify macroscopicity of quantum states \cite{Froewis12,Froewis15,Jeong15,Bjork04,Korsbakken07,Marquardt08,Sekatski14,Shimizu02,Shimizu05,Lee11,Nimmrichter13}. The literature about this topic is diverse and various measures are mutually compared in Refs.\ \cite{Froewis12,Froewis15} and summarized in Ref.\  \cite{Jeong15}. Generally speaking, a macroscopic quantum state (MQS) is a state  capable of displaying macroscopic quantum effects that can be utilized to validate quantum mechanics (against classical theories) on a macroscopic scale. An important task is the identification of a characteristic parameter that measures the ``size'' or ``macroscopicity'' of a certain quantum state~\cite{Leggett02}, such as the characteristic energy, mass, number of elementary constituents, etc. Here we focus on the case of macroscopically large number of particles $N$ that interact via a local Hamiltonian.

\begin{figure}[t]
\centering
\includegraphics[width=0.72\linewidth]{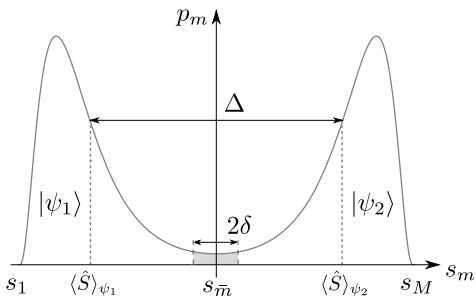}
\caption{The distribution $p_m$ of eigenvalues $s_m$ of an additive observable $\hat{S}$ for a MS state $\ket{\psi_1}+\ket{\psi_2}$. A continuous curve is used for aesthetic purposes. The distribution has two well-resolved regions (left and right from the separation point $s_{\bar m}$) each corresponding to the superimposed semiclassical states $\ket{\psi_1}$ and $\ket{\psi_2}$, respectively. The distance between the regions is $\Delta:=|\avg{\hat S}_{\psi_2}-\avg{\hat S}_{\psi_1}|$. The separation probability related to the finite-sized shaded segment $|s-s_{\bar m}|\leq\delta=O(N^0)$ should be vanishing in the macroscopic limit $N\to\infty$.}
\label{fig:1}
\end{figure}

An important subclass of MQS are macroscopic superpositions (MS): states of the type $\ket{\psi}=\ket{\psi_1}+\ket{\psi_2}$, where $\ket{\psi_{1,2}}$ are macroscopically well-distinguishable states. However, such a definition is not operational as there are infinitely many decompositions of the kind $\ket{\psi_1}+\ket{\psi_2}$ and it might not be clear how to unambiguously identify the ``semiclassical'' components of the MS. Therefore, we define MS with respect to a measurement of an additive (collective) observable \cite{Froewis12,Froewis15,Bjork04,Shimizu02,Shimizu05}. A pure state $\ket{\psi}$ is MS if a measurement of some additive observable $\hat S$ can sharply distinguish the semiclassical states that constitute MS; e.g., the distribution of eigenvalues of $\hat S$ exhibits two well-resolvable regions (see Fig.\ \ref{fig:1}). Our main focus here is on {\it (i) the possibility of the natural appearance of such states as unique ground states of macroscopic quantum systems} and, consequently, {\it (ii) the feasibility of preparing MS by simply cooling down such systems}. The latter might be achievable provided that the system has a unique MS ground state; i.e., there is a finite energy gap in the thermodynamic limit. In this respect, it was proven that no MS of ``locally distinguishable'' states can be the unique ground state of $N$ spins described by a local Hamiltonian whose energy gap is at least $O(1/{\rm poly}(N))$ \cite{Froewis13}. Conversely, numerical evidence was given in Ref.\ \cite{Morimae10} that the energy gap of a certain $N$-qubit Hamiltonian decays exponentially fast in the macroscopic limit when its ground state actually is MS. Moreover, relation between the spectral gap and ground state properties of spin lattice systems was studied in Refs.\ \cite{Kuwahara15,Kuwahara16}.

We provide a simple sufficient criterion enforcing the energy gap to vanish in the thermodynamic limit for a very general class of ground states of local many-body Hamiltonians. The most important feature of our approach is an operational method to identify semiclassical states that constitute the macroscopic superposition. We show that in many cases local Hamiltonians are not capable of linking such states, so that the corresponding MS can only represent a degenerate ground state in the macroscopic limit. Our main theorem provides an interesting relation between the energy gap and the order of interaction (i.e., the number $K$ in the case of a $K$-body interaction). Therefore, one may derive the lowest order of interaction for which a given MS might be a unique ground state. We discuss our results in the context of different physical systems and various proposals for preparation of MS. Furthermore, we show that a certain class of states that are not even considered to be macroscopically quantum (e.g., $W$ states) cannot naturally exist as ground states of gapped local Hamiltonians. Finally, we demonstrate that the methods and results derived here are relevant for quantum marginal related problems.

\emph{Preliminaries.}---Let us consider a system of $N$ interacting particles described by a $K$-local Hamiltonian $\hat
H=\sum_{(i_1,i_2,\dots,i_K)\in {\cal I}_N^{(K)}}\hat H_{i_1 i_2\dots i_K}^{}\vphantom{\frac{1}{\sum}}$, where $\hat H_{i_1 i_2 \dots i_K}^{}$ is the contribution due to interaction between particles $i_1,i_2,\dots,i_K$ and ${\cal I}_N^{(K)}$ is the set of all $K$-tuples of $N$ interacting particles. We call $K$ the order of interaction. For instance, usual physical interactions are pairwise with the order $K=2$.

We begin with the following general lemma:

\emph{Lemma.}---Let a Hamiltonian $\hat H$ have a unique ground state of the form $\ket{\psi}=a_1\ket{\psi_1}+
a_2\ket{\psi_2}$, where $\ket{\psi_{1,2}}$ are normalized, $\braket{\psi_2}{\psi_1}=\lambda$ and  $a_1,a_2>0$. Then the energy gap
$\Delta E$ satisfies the inequality\vspace{-1mm}
\begin{align}\label{eq:gap}
\Delta E\leq\frac{|\bra{\psi_2}\hat H\ket{\psi_1}-\lambda E_0|}{a_1 a_2(1-|\lambda|^2)},
\end{align}\vspace{-4mm}\\\noindent
where $E_0$ denotes the ground state energy (see Supplemental Material \cite{SM} for the proof).

Without loss of generality we set $E_0=0$ hereafter. We start our analysis with the simple observation that the energy gap is essentially upper bounded by a magnitude of the matrix element $\bra{\psi_2}\hat H\ket{\psi_1}=H_{21}$ [assuming that the overlap $\lambda$ is vanishingly small and $a_{1,2}=O(N^0)$ when $N\to\infty$]. Therefore, the system cannot have a finite gap in the macroscopic limit if $H_{21}$ is vanishing when $N\to\infty$.

An archetypal example of MS is a so-called GHZ state \cite{GHZ}, closely related to an original Schr\"odinger's proposal as it is a superposition of two macroscopically distinct states of $N$ particles, i.e., $\ket{\psi}\propto\ket{\varphi_1}^{\ox\!N}+ \ket{\varphi_2}^{\ox\!N}$. The states $\ket{\varphi_{1,2}}$ are normalized with the fixed nonzero overlap $\omega= |\braket{\varphi_1}{\varphi_2}|<1$. Here, one can naturally identify the two constituents $\ket{\psi_{1,2}}=\ket{\varphi_{1,2}}^{\ox\!N}$ with exponentially small overlap $|\lambda|=\omega^N$ and $a_{1,2}\xrightarrow{N\to\infty}1/\sqrt{2}$. Denote by $H^{[K]}_{21}$ the maximal magnitude of all matrix elements $\bra{\varphi_2}^{\otimes K}\hat H_{i_1,i_2,\dots, i_K}\ket{\varphi_1}^{\otimes K}$.  The value of $H^{[K]}_{21}$ does not scale with $N$ and solely depends on the nature of the interaction. It is not difficult to see that
\begin{eqnarray}
|H_{21}|&\leq&|\mbox{\;\!}{\cal I}_N^{(K)}|\;\!\omega^{N-K}H^{[K]}_{21}\leq\binom{N}{K}\;\!\omega^{N-K}H^{[K]}_{21},
\end{eqnarray}
since for $K$ fixed the total number of interaction terms grows at most polynomially with $N$, i.e., $|\mbox{\;\!}{\cal I}_N^{(K)}|\leq \binom{N}{K}=O(N^K)$. Therefore, we conclude that the energy gap vanishes exponentially fast when $N\to\infty$, as long as the order of interaction is fixed. In other words, all the states $\ket{\psi(\alpha)}\propto\ket{\varphi_1}^{\ox\!N}+e^{i\alpha} \ket{\varphi_2}^{\ox\!N}$ give the same energy in the thermodynamic limit and the ground state becomes at least doubly degenerate. Consequently, cooling down the system towards zero temperature will result in a classical mixture $\frac{1}{2}\ketbra{\psi(0)}{\psi(0)} +\frac{1}{2}\ketbra{\psi(\pi)}{\psi(\pi)}$. In order to make the energy gap finite in the thermodynamic limit, it is necessary that the order of interaction $K$ grows with the number of particles $N$, which is usually considered nonphysical. This reasoning can be trivially extended to a finite sum $\ket{\varphi_1}^{\ox\!N}+\ldots+\ket{\varphi_n}^{\ox\!N}$ of macroscopically distinguishable states, i.e., $\braket{\varphi_i}{\varphi_j} =O(N^0)$ when $i\neq j$. In the Supplemental Material \cite{SM} we show that the same result holds for a  more general class of states, i.e., the superpositions of locally distinguishable states that have been considered in literature as a natural generalization of the GHZ-like states \cite{Froewis12,Korsbakken07}.

Whereas the previous examples are fairly easy to grasp, as the superimposed states are identifiable by definition, such a
clean prescription is not {\it a priori} available for arbitrary MQS. Therefore, we continue our analysis by invoking a
measurement of some collective observable $\hat S$ that should serve as a reference point to identify $\ket{\psi_{1,2}}$.

Consider a system of $N$ particles in a total Hilbert space $\mathcal{H}^{N}=\otimes^{N}_{i=1}\mathcal{H}_i$, with $\mathrm{dim} (\mathcal{H}_i)=d$. Let $\hat S=\sum_{i=1}^N\hat S_i$ be an additive observable. The single-particle operators satisfy $\hat S_i \ket{\sigma_i,\mu_i}_i=\sigma_i\ket{\sigma_i,\mu_i}_i$, where $\sigma_i\in\{\varsigma_1<\varsigma_2<\ldots<\varsigma_\ell\}$ and $2\le\ell\le d$, while $\mu_i=1,\ldots,\mu(\sigma_i)$ enumerate the degeneracies obeying $\sum_{l=1}^\ell \mu(\varsigma_l)=d$. We denote the different eigenvalues of $\hat{S}$ by $s_1<s_2<\!\ldots\!<s_M$, where $s_m=\sum_{l=1}^\ell n_{m,l}\:\!\varsigma_l$, $n_{m,l}\in\mathbb{N}_0$ and $\sum_{l=1}^\ell n_{m,l}=N$. Clearly, $s_1=N\varsigma_1$ and $s_M=N\varsigma_\ell$. The states $\ket{\bm{\sigma},\bm{\mu}}= \ox_{i=1}^N\ket{\sigma_i,\mu_i}_i$ constitute a complete basis in $\mathcal{H}^{N}$, i.e., $\sum_{\bm{\sigma}}\sum_{\bm{\mu}} \ket{\bm{\sigma},\bm{\mu}}\bra{\bm{\sigma},\bm{\mu}} =\openone$, where $\bm{\sigma}=(\sigma_1,\sigma_2,\ldots,\sigma_N)$ and $\bm{\mu}=(\mu_1,\mu_2,\ldots,\mu_N)$. This yields a decomposition\vspace{-1mm}
\begin{equation}
\ket{\psi}=\sum_{\bm{\sigma}}\sum_{\bm{\mu}} \ket{\bm{\sigma},\bm{\mu}}\braket{\bm{\sigma},\bm{\mu}}{\psi}=\sum_{m=1}^M\sqrt{p_m}\ket{s_m},\vspace{-1mm}
\end{equation}
where $\hat S\ket{s_m}=s_m\ket{s_m}$, and $\ket{s_m}$ contains all the terms from the multisums such that $\sum_{i=1}^N\sigma_i=s_m$. The numbers $p_m\geq 0$ correspond to the probabilities of obtaining the value $s_m$ when measuring the observable $\hat S$ in the state $\ket{\psi}$, hence, $\sum_{m=1}^M p_m=1$.

Now, if the state $\ket{\psi}$ is MS of two states $\ket{\psi_1}$ and $\ket{\psi_2}$, then we expect that the probability distribution ${\cal P}_\psi=\{p_m\}_{m=1}^M$ has two distinguishable regions with corresponding probabilities of the order $O(N^0)$ and with vanishingly small probability within the finite-sized bordering segment around some eigenvalue $s_{\bar m}$ of $\hat S$ (see Fig.\ \ref{fig:1}). Those regions should precisely be related to the semiclassical constituents of the state $\ket{\psi}$. The distance between the regions $\Delta:=|\avg{\hat S}_{\psi_2}-\avg{\hat S}_{\psi_1}|$ is closely related to the fluctuation of the observable $\hat S$ in the state $\ket{\psi}$ and it is commonly assumed that MS displays $\Delta=O(N)$~\cite{Froewis12,Shimizu05,Lee11}. However, we will address quantum states from another aspect, which will render our main result independent of $\Delta$. Namely, the prime quantity in our analysis is {\it the separation probability} $P_\psi(|s-s_{\bar m}|\leq\delta)$, i.e., the probability of finding the result $s$, when measuring $\hat S$, within a tiny segment of size $2\delta=O(N^0)$ centered at {\it the separation point} $s_{\bar m}$. We will provide an upper bound on the energy gap, which essentially depends on the separation probability and the order of interaction. Thus, the interplay between the two will have a crucial role in vanishing of the gap.

Next, we will make use of $s_{\bar m}$ to express the ground state in the form of a superposition
\begin{equation}\label{eq:psi}
\ket{\psi}=a_1\ket{\psi_1}+a_2\ket{\psi_2},
\end{equation}
with
\begin{align}
a_1\ket{\psi_1}=\sum_{m=1}^{\bar{m}-1}\sqrt{p_m}\ket{s_m},\quad
a_2\ket{\psi_2}=\sum_{m=\bar{m}}^M\sqrt{p_m}\ket{s_m},
\end{align}
where $a_1=(p_1+\ldots+p_{\bar{m}-1})^{1/2}$, $a_2=(p_{\bar m}+\ldots+p_M)^{1/2}$, and, presumably, $a_{1,2}=O(N^0)$. By construction, one has $\braket{\psi_2}{\psi_1}=0$. We will employ the introduced separation to derive an upper estimate of the energy gap.

Let us suppose that the Hamiltonian of the physical system is 2-local, i.e., $\hat H=\sum_{(i,j)\in {\cal I}_N^{(2)}}\hat H_{ij}^{}\vphantom{\frac{1}{\sum}}$, where $\hat H_{ij}^{}$ represents pairwise interaction between particles $i$ and $j$ and ${\cal I}_N^{(2)}$ is the set of pairs of interacting particles. Obviously, the number of interaction terms in the Hamiltonian satisfies $|\mbox{\;\!}{\cal I}_N^{(2)}|\leq N(N\!-\!1)/2=O(N^2)$. The magnitude of the matrix element in the inequality (\ref{eq:gap}) can be estimated in order to obtain the following central result:

\emph{Theorem.}---Under the assumptions given in the text, the energy gap of the system is bounded as\vspace{-1mm}
\begin{equation}\label{eq:Gap}
\Delta E\le\frac{|\mbox{\;\!}{\cal I}_N^{(2)}|}{2\:\!a_1^2 a_2^2}\max_{(i,j)\in{\cal I}_N^{(2)}}\|\hat H_{ij}^{}\|\cdot P_\psi(|s-s_{\bar m}|\le 2\delta_\varsigma),
\end{equation}\vspace{-2mm}\\\noindent
where $\max_{(i,j)\in{\cal I}_N^{(2)}}\|\hat H_{ij}^{}\|$ sets the characteristic energy scale (independent of $N$) and $\delta_\varsigma=\varsigma_\ell-\varsigma_1$. Here, $\|\mbox{\;\!}\cdot\mbox{\;\!}\|$ denotes the operator spectral norm. The complete proof is given in the Supplemental Material \cite{SM}.

The bound (\ref{eq:Gap}) is valid for any $s_{\bar m}$, which has been arbitrary up to now. Clearly, one should select $\hat S$ and the corresponding $s_{\bar m}$ so that $P_\psi(|s-s_{\bar m}|\leq 2\delta_\varsigma)$ vanishes as fast as possible for $N\to\infty$. In the previously discussed GHZ-like case, the separation probability scales as $\exp[-O(N)]$ and the energy gap vanishes exponentially fast with $N$. Furthermore, it is clear that for any state exhibiting $P_{\psi}=o(1/N^2)$ the gap will vanish in the thermodynamic limit and the state can only represent a degenerate ground state. In general, such a state does not necessarily display anomalous fluctuation of $\hat S$. One can even find examples where $\Delta=O(N^0)$ [such as $\ket{\psi}=(\ket{s_{m_1}}+\ket{s_{m_2}})/\sqrt 2$, where $s_{m_1}=s_{\bar m}-2\delta$ and $s_{m_2}=s_{\bar m}+2\delta$, with $\delta>\delta_\varsigma$]. Conversely, when the system features a finite energy gap, the relation (\ref{eq:Gap}) puts a lower bound $P_\psi(|s-s_{\bar m}|\leq 2\delta_\varsigma) \ge O(1/N^2)$ for any gapped $2$-local Hamiltonian and arbitrary observable $\hat S$.

The appearance of probabilities corresponding to the interval of size $4\delta_\varsigma$ centered at $s_{\bar m}$ is a direct consequence of the 2-local nature of the Hamiltonian. We note that the Theorem could easily be generalized for arbitrary $K$-local Hamiltonians. In that case, one would consider the set ${\cal I}_N^{(K)}\!$ of $K$-tuples of interacting particles, for which $|\mbox{\:\!}{\cal I}_N^{(K)}|\leq\binom{N}{K}=O(N^K)$, and the corresponding estimate of the gap would involve the probability $P_\psi(|s-s_{\bar m}|\leq 2K\delta_\varsigma)$. Thus, for a gapped $K$-local Hamiltonian we conclude that the best possible separation probability one can achieve for a ground state is asymptotically lower bounded by $O(1/N^K)$. Consequently, all the states exhibiting the scaling $P_{\psi}=o(1/N^K)$ are excluded as possible unique ground states.

\emph{Various examples.}---Our general result nicely complies with the investigation of ground states of various physical systems. For example, a twofold fragmented condensate of interacting bosons trapped in a single well \cite{Fischer09} features a doubly degenerate ground state, in the thermodynamic limit. It was shown in Ref.\ \cite{Fischer15} that in the appropriate Fock space basis the corresponding ground states are identical to the photon cat states. In accordance with our findings, the proposed preparation of these states requires other means than simple cooling, i.e., the rapid sweep of interaction couplings \cite{Fischer11}. Another example is a one-dimensional array of circuit quantum electrodynamic (cQED) systems in the ultrastrong cavity-qubit coupling regime \cite{Hwang13}. The authors showed that the photon hopping between cavities can be mapped to the Ising interaction between the lowest two levels of individual cQED of the chain. Based on the mapping, they found two nearly degenerate GHZ-type ground states with energy splitting exponentially small in the system size. Again, this is in perfect agreement with our results. Moreover, we mention the study of a bosonic Josephson junction made of $N$ ultracold and dilute atoms confined by a quasi-one-dimensional double-well potential within the two-site Bose-Hubbard model framework \cite{Mazzarella11}. Detailed treatment showed that the ground state of the system evolves towards NOON state when increasing attractive interatomic interaction. The estimated gap between two lowest energy states vanishes exponentially with $N$, in full compliance with our considerations. Our work also nicely agrees with Ref.\ \cite{Huang06} where the possibility of creating many-particle catlike states was examined for a Bose-Einstein condensate trapped in a double-well potential. It was discussed in detail that creating cat states via adiabatic manipulation of the many-body ground state is experimentally unfeasible due to the fact that the end state is nearly degenerate with the first-excited state; hence, such a process would require an exponentially long time. This difficulty was surpassed by proposing to exploit dynamic evolution following a sudden flipping of the sign of the atomic interaction, accomplished via Feshbach resonance technique \cite{Feshbach}. Finally, we mention that our treatment assumes a close correspondence between the macroscopicity of the system and the number of its constituents. However, the macroscopicity might be related to other quantities and only weakly depend on the system size. SQUID systems, which were proposed as good candidates to host the ``genuine'' MS \cite{Leggett02}, are a paramount example of that. Although our results are not directly applicable to such a case, in the Supplemental Material \cite{SM} we provide a discussion of SQUIDs showing some similarities with our findings.

Our generic analysis demonstrates that more sophisticated experimental techniques are needed for the preparation of a variety of macroscopic superpositions in the thermodynamic limit. This may require some form of dynamical driving of a system, as in the mentioned examples, advanced matter-wave interferometric approaches \cite{Doerre14} or use of demanding postselection techniques \cite{Pan16}.

Furthermore, we present an example to demonstrate that our results can be used to address the states that are more general than MQS (see Supplemental Material \cite{SM}). Consider a lattice model of $N$ spin-$1/2$ particles interacting with the fixed number of neighbors. Thus, one has $d=2$, $\ell=2$, $\delta_\varsigma=1$, and $|\mbox{\;\!}{\cal I}_N^{(2)}|=O(N)$. In order to prove that the model becomes gapless in the limit $N\to\infty$, one has to find an appropriate additive observable $\hat S$ for which the ground-state-related separation probability vanishes as $o(1/N)$. Collective states that naturally appear in spin systems are the Dicke states \cite{Dicke54} $\ket{j,m}$ ($m=-j,\ldots,j$), where $j=N/2$. They are permutation invariant and satisfy $\hat J^2\ket{j,m}=j(j+1)\ket{j,m}$ and $\hat J_z\ket{j,m}=m\ket{j,m}$. All Dicke states are unique ground states of some fully 2-local, gapped Hamiltonian for which $|\mbox{\;\!}{\cal I}_N^{(2)}|=N(N-1)/2$ (all the particles mutually interact pair wisely, such as indistinguishable particles) \cite{Froewis13}. However, such Hamiltonians do not correspond to the present case. Therefore, we will show that, for example, an $N$-qubit $W$ state $\ket{j,j-1}$, which represents the case of symmetrically distributed one-spin excitations, cannot be a unique ground state of any considered spin-lattice model. First, we find the appropriate collective observable to be $\hat J_x$. Let $\ket{j,m}_x$ ($m=-j,\ldots,j$) be the common eigenbasis of $\hat J^2$ and $\hat J_x$. The related probability distribution is $p_m=|\braket{j,j-1}{j,m}_x|^2$ (see Fig.\ 1 in the Supplemental Material \cite{SM}), $s_m=m$, and we choose $s_{\bar m}=0$ for $j$ integer or $s_{\bar m}=1/2$ for $j$ half-integer. As presented in the Supplemental Material \cite{SM}, we find
\begin{align}
p_m=\frac{2m^2}{2^{2j}j}\binom{2j}{j+m}\sim\frac{2m^2}{\sqrt{\pi}j^{3/2}},
\end{align}
where the last asymptotic behavior holds for fixed $m$ and $j\to\infty$. We conclude that the separation probability $P_\psi(|s-s_{\bar m}|\leq 2)$ scales as $O(1/j^{3/2})$, i.e., $O(1/N^{3/2})$. Thus, the $W$ state can only be a degenerate ground state of the arbitrary spin-lattice model considered here. Moreover, the distance between the two peaks has sublinear asymptotic scaling $\sim\sqrt{2N}$. Hence, the $W$ state is an example of a state that is not even a MQS according to the anomalous fluctuation criterion, but is nevertheless amenable to our present analysis.

Finally, our results can be naturally related to quantum marginal problem \cite{Klyachko,Coleman}. There, the main task is to check whether or not a given set of marginal states $\hat{\bm{\rho}}=(\hat\rho_{s_1},\hat\rho_{s_2},\dots)$ can be extended to some $N$-particle quantum state $\hat\varrho^{[N]}$, i.e., $\hat\rho_{s_k}=\mathrm{Tr}_{\,\overline{\!s_k\!}} \,\hat\varrho^{[N]}$, where $s_k$ denotes a subset of $N$ particles. The set of all representable marginals $\hat{\bm{\rho}}$ is convex and completely characterized by its extremal points (for finite-dimensional systems); therefore, their identification is of great importance. On the other hand, the set of extremal points is in unique correspondence to the set of $N$-particle nondegenerate ground states of the local Hamiltonians~\cite{Coleman}. Namely, for a given Hamiltonian $\hat H=\sum_{k}\hat H_{s_k}$, where $\hat H_{s_k}$ denotes local Hamiltonian acting on the subset of particles $s_k$, we have $E=\mathrm{Tr}(\hat\varrho^{[N]}\hat H)=\sum_k\mathrm{Tr}_{s_k}(\hat\rho_{s_k}\hat H_{s_k})= \mathrm{Tr}(\hat{\bm{\rho}} \hat{\bm{H}})$, where $\hat{\bm{H}}=(\hat H_{s_1},\hat H_{s_2},\dots)$. Thus, the energy $E$ is a linear functional on the set of all representable marginals $\hat{\bm{\rho}}$ and it reaches its extreme values on the set of nondegenerate ground states. Our criterion (\ref{eq:Gap}) implies that a large class of degenerate ground states (in the thermodynamic limit) has the set of marginals that cannot be extremal.

\emph{Summary and outlook.}---In this Letter we provided a powerful generic method to analyze the possibility for ground states of gapped many-body quantum systems to be superpositions of macroscopically distinct quantum states. We have ruled out a large class of quantum states that cannot be prepared by simply cooling macroscopic quantum systems that exhibit interactions involving some finite number of their constituents. For such a state, we require that the separation probability, related to the small segment around the separation point between its two semiclassical components, vanishes sufficiently fast in the thermodynamic limit. We expect our results to be valuable for future experiments aiming at preparing quantum states that exhibit macroscopic quantum properties. Furthermore, we have shown that our study is relevant for quantum marginal problem.

\begin{acknowledgments}
We thank \v{C}aslav Brukner, Nikola Paunkovi\'c, and Jacques Pienaar for helpful comments and gratefully acknowledge financial support from the European Commission through the projects RAQUEL (No.\ 323970) and QUCHIP (No.\ 641039).
\end{acknowledgments}%


\setcounter{equation}{0}
\setcounter{figure}{0}

\section{Supplemental Material}

In this Supplemental Material we provide the proofs of the statements from the main text. Furthermore, we study an example of certain superpositions of Dicke states as possible ground states of local Hamiltonians of spin-lattice models. At the end, we provide the discussion about quantum superpositions in superconducting quantum interference devices (SQUIDs).

\subsection{Proof of the Lemma}

To ease the notation, we introduce the operator $\hat h=\hat H-E_0$. The ground state energy of $\hat{h}$ is zero, whereas the energy of the first excited state is equal to the energy gap $\Delta E$ of $\hat{H}$. The ground state $\ket{\psi}=a_1\ket{\psi_1}+a_2\ket{\psi_2}$ satisfies the condition $\hat h\ket{\psi}=0$. Therefore, we get the following set of equations
\begin{subequations}\label{eq:hsys}
\begin{align}
a_1 h_{11}+a_2 h_{12}&=0,\\
a_1 h_{21}+a_2 h_{22}&=0,
\end{align}
\end{subequations}
with $h_{ij}=\bra{\psi_i}\hat h\ket{\psi_j}$. The linear system above has non-trivial solutions if its determinant is zero. Hence, $h_{11}h_{22}=h_{12}h_{21}\equiv |h_{21}|^2$. Obviously, $h_{11},h_{22}\ge 0$ as the ground state energy of $\hat{h}$ is zero. Consider now the expansion $\ket{\psi_1}=c_1\ket{\psi}+\sqrt{1-|c_1|^2}\ket{\psi^\perp}$, where $\ket{\psi^{\perp}}$ is the linear combination of $\ket{\psi_1}$ and $\ket{\psi_2}$ such that $\braket{\psi}{\psi^\perp}=0$. We get the following inequality
\begin{align}
\bra{\psi_1}\hat h\ket{\psi_1}=(1-|c_1|^2)\bra{\psi^\perp}\hat h\ket{\psi^\perp}\geq (1-|c_1|^2)\Delta E.
\end{align}
The last inequality follows from the fact that the lowest energy of $\hat h$ in the subspace orthogonal to $\ket{\psi}$ is $\Delta E$. It is easy to obtain $1-|c_1|^2=|a_2|^2(1-|\lambda|^2)$, so that
\begin{align}
\Delta E\leq\frac{\bra{\psi_1}\hat h\ket{\psi_1}}{1-|c_1|^2}=\frac{h_{11}}{|a_2|^2(1-|\lambda|^2)}.
\end{align}
From the equations (\ref{eq:hsys}) we find $h_{11}/|a_2|^2=|h_{12}|/|a_1 a_2|=|h_{21}|/|a_1 a_2|=h_{22}/|a_1|^2$. Consequently, we get
\begin{align}
\Delta E\leq\frac{|h_{21}|}{|a_1 a_2|(1-|\lambda|^2)}.
\end{align}
Recalling the assumptions $a_1,a_2>0$ from the main text, the previous result proves the Lemma.

\subsection{Macroscopic superpositions of ``locally distinguishable'' states}

Let us assume that the unique ground state $\ket{\psi}$ is a macroscopic superposition of two states $\ket{\psi_1}$ and $\ket{\psi_2}$. We will rely on the measurement-based measure of the size of macroscopic quantum superpositions in terms of ``local distinguishability'', introduced in Ref.\ \cite{SM_Korsbakken07} and elaborated in Ref.\ \cite{SM_Froewis12}. In this context, it was shown in Ref.\ \cite{SM_Froewis13} that if the energy gap scales as $O(1/{\rm poly}(N))$, then no MS of locally distinguishable states can be the unique ground state of $N$ spins described by a local Hamiltonian. Our goal here is to prove the opposite $-$ if such a state is a ground state of local Hamiltonian, then the energy gap vanishes exponentially fast in the macroscopic limit. Following \cite{SM_Froewis12,SM_Korsbakken07}, we divide $N$ particles into a maximal number $\tilde N$ of distinct groups of particles such that $\ket{\psi_1}$ can be distinguished from $\ket{\psi_2}$ with probability $P>1/2$ by performing a measurement on any single group. The superposition state $\ket{\psi}$ is called macroscopic if $\tilde N=O(N)$. To avoid cumbersome notation, we assume that the size of every group is $N/\tilde{N}$ and introduce the abbreviation $[N]=\{1,2,\ldots,N\}$. In the sequel, we  will derive an exponential bound for the magnitude of the matrix element of a 2-local Hamiltonian
\begin{align}\label{eq:h12}
\bra{\psi_2}\hat H\ket{\psi_1}&=\sum_{(i,j)\in{\cal I}_N^{(2)}}\bra{\psi_2}\hat H_{ij}^{}\ket{\psi_1}\nonumber\\
&=\sum_{(i,j)\in{\cal I}_N^{(2)}} \Tr{i,j}{\hat H_{ij}^{}\Tr{[N]\setminus\{i,j\}}{\ketbra{\psi_1}{\psi_2}}}[\Big].
\end{align}\vspace{-3mm}

Our approach is based on the one given in the Appendix C of Ref.\ \cite{SM_Froewis13}. Denote by $\hat A^{(k)}$ the measurement operator on group $k$ that optimally distinguishes the states $\ket{\psi_1}$ and $\ket{\psi_2}$. One can always choose it so that its spectrum is $\{-1,+1\}$. In such a case, the success probability to distinguish the two states is given by $P=1/2+1/4\,|\avg{\hat A^{(k)}}_{\psi_1}-\avg{\hat A^{(k)}}_{\psi_2}|$. Since $P>1/2$, we can assume that $-1\leq \avg{\hat A^{(k)}}_{\psi_2}<\avg{\hat A^{(k)}}_{\psi_1}\leq 1$ for all $k$. For the $k$th group, the projection operator on outcome $\alpha$ is denoted by $\hat\Pi^{(k)}_{\alpha\vphantom{l}}$. The measurement probabilities are then $\|\hat\Pi^{(k)}_{\pm 1\vphantom{l}}\ket{\psi_i}\|^2=(1\pm \avg{\hat A^{(k)}}_{\psi_i})/2\equiv p_{i,\pm}^{}$ for $i=1,2$. One additional comment is in order. Namely, in the generic case $\tilde N$ can depend on the success probability $P$. As discussed in Ref.\ \cite{SM_Froewis12}, the additional assumption, that the measurements on any group do not influence the measurement outcomes on other groups, resolves this issue. It basically means that only correlations within the groups exist and not among different groups. Thus, we formally require that $\avg{\hat A^{(k)}\hat A^{(k')}}_{\psi_i}=\avg{\hat A^{(k)}}_{\psi_i}\avg{\hat A^{(k')}}_{\psi_i}$, for $i=1,2$ and for all groups $k\neq k'$. This in turn implies the factorization $\|\hat\Pi^{(k)}_{\alpha\vphantom{l}} \hat\Pi^{(k')}_{\alpha'\vphantom{l}}\ket{\psi_i}\|^2=p^{}_{i,\alpha\vphantom{l}}\, p^{}_{i,\alpha'\vphantom{l}}$ for $i=1,2$. Similar factorization is found for the joint probabilities of the results of measurements on more than two distinct groups.

Denote by ${\bm{\Gamma}}= \{\Gamma^{(k)}\}_{k=1}^{\tilde N}$ the set of the considered distinct groups of particles. Let us examine the partial trace of $\ketbra{\psi_1}{\psi_2}$ over $n$ groups $\{\Gamma^{(k_1)},\ldots,\Gamma^{(k_n)}\}\equiv {\bm{\Gamma}}^{(\bm{k})}_n$
\begin{widetext}\vspace{-4mm}
\begin{align}\label{eq:pTr}
\Tr{{\bm{\Gamma}}^{(\bm{k})}_n}{\ketbra{\psi_1}{\psi_2}}&=\Tr{{\bm{\Gamma}}^{(\bm{k})}_n}{\ox_{i=1}^n(\hat\Pi^{(k_i)}_{+1}+ \hat\Pi^{(k_i)}_{-1})\,\ketbra{\psi_1}{\psi_2}\ox_{j=1}^n(\hat\Pi^{(k_j)}_{+1}+\hat\Pi^{(k_j)}_{-1})}[\Big]\nonumber\\
&=\Tr{{\bm{\Gamma}}^{(\bm{k})}_n}{\sum_{\alpha_1,\ldots,\alpha_n=\pm 1}\sum_{\alpha_1',\ldots,\alpha_n'=\pm 1}
\hat\Pi^{(k_1)}_{\alpha_1\vphantom{l}}\ldots\hat\Pi^{(k_n)}_{\alpha_n\vphantom{l}}\,\ketbra{\psi_1}{\psi_2}\,
\hat\Pi^{(k_1)}_{\alpha_1'\vphantom{l}}\ldots\hat\Pi^{(k_n)}_{\alpha_n'\vphantom{l}}}[\Big]\nonumber\\
&=\Tr{{\bm{\Gamma}}^{(\bm{k})}_n}{\sum_{\alpha_1,\ldots,\alpha_n=\pm 1}\hat\Pi^{(k_1)}_{\alpha_1\vphantom{l}} \ldots\hat\Pi^{(k_n)}_{\alpha_n\vphantom{l}}\,\ketbra{\psi_1}{\psi_2}\,\hat\Pi^{(k_1)}_{\alpha_1\vphantom{l}}\ldots \hat\Pi^{(k_n)}_{\alpha_n\vphantom{l}}}[\Big]\nonumber\\
&=\Tr{{\bm{\Gamma}}^{(\bm{k})}_n}{\sum_{\bm{\alpha}\in\{-1,1\}^n}\hat\Pi^{(\bm{k})}_{\bm{\alpha}\vphantom{l}}\,\ketbra{\psi_1}{\psi_2}\, \hat\Pi^{(\bm{k})}_{\bm{\alpha}\vphantom{l}}}[\Big],
\end{align}\vspace{-3mm}
\end{widetext}
where $\hat\Pi^{(\bm{k})}_{\bm{\alpha}\vphantom{l}}=\hat\Pi^{(k_1)}_{\alpha_1\vphantom{l}}\ldots\hat\Pi^{(k_n)}_{\alpha_n\vphantom{l}}$. We used the orthogonality $\hat\Pi^{(k)}_{\alpha\vphantom{l}}\hat\Pi^{(k)}_{\alpha'\vphantom{l}}=\delta_{\alpha\alpha'}^{} \hat\Pi^{(k)}_{\alpha\vphantom{l}}$ and the completeness relation $\hat\Pi^{(k)}_{+1\vphantom{l}}+\hat\Pi^{(k)}_{-1\vphantom{l}}= \mathbb{I}^{(k)}$. In evaluating (\ref{eq:h12}) we will encounter two types of terms, the ones where both particles $i$ and $j$ belong to two different groups, say $\Gamma^{(k_i)}$ and $\Gamma^{(k_j)}$, and the ones where they belong to the same group, say $\Gamma^{(k_{ij})}$. In both cases we have to perform partial traces over at least $\tilde N-2$ groups of particles. To make the derivation more compact, we will treat the two cases on the same footing. Namely, we shall introduce ${\bm{\Gamma}_{ij}}=\{\Gamma^{(k_i)},\Gamma^{(k_j)}\}$ in the former case. In the latter case, we define ${\bm{\Gamma}_{ij}}=\{\Gamma^{(k_{ij})},\Gamma^{(k_{ij}^*)}\}$, where to each group $k$ we assign its ``partner'' group $k^*\neq k$. For instance, one may set $1^*=2$, $2^*=3,\ldots$, $(N-1)^*=N$ and $N^*=1$. We then have\vspace{-2mm}
\begin{align}
\Tr{[N]\setminus\{i,j\}}{\ketbra{\psi_1}{\psi_2}}={\rm Tr}_{\bm{\Gamma}_{ij}\setminus\{i,j\}}{\Tr{{\bm{\Gamma}} \setminus\bm{\Gamma}_{ij}}{\ketbra{\psi_1}{\psi_2}}},
\end{align}\vspace{-5mm}\\\noindent
where the last partial trace is always over $\tilde N-2$ groups. Let $\bm{k}_{i,j}$ label those groups. Combining the previous, we get
\begin{widetext}\vspace{-5mm}
\begin{align}
|\bra{\psi_2}\hat H\ket{\psi_1}|&=\Big|\sum_{(i,j)\in{\cal I}_N^{(2)}} \Tr{i,j}{\hat H_{ij}^{}{\rm Tr}_{\bm{\Gamma}_{ij}\setminus\{i,j\}}{\Tr{\bm{\Gamma}\setminus\bm{\Gamma}_{ij}}{\ketbra{\psi_1}{\psi_2}}}}[\Big]\Big|
\nonumber\displaybreak[1]\\
&=\Big|\sum_{(i,j)\in{\cal I}_N^{(2)}} \Tr{i,j}{\hat H_{ij}^{}{\rm Tr}_{\bm{\Gamma}_{ij}\setminus\{i,j\}}{\Tr{\bm{\Gamma}\setminus \bm{\Gamma}_{ij}}{\!\!\sum_{\bm{\alpha}\in\{-1,1\}^{\tilde N-2}}\!\!\hat\Pi^{(\bm{k}_{i,j})}_{\bm{\alpha}\vphantom{l}}\,\ketbra{\psi_1}{\psi_2} \,\hat\Pi^{(\bm{k}_{i,j})}_{\bm{\alpha}\vphantom{l}}}}}[\Big]\Big|\nonumber\displaybreak[1]\\
&=\Big|\sum_{(i,j)\in{\cal I}_N^{(2)}} \sum_{\bm{\alpha}\in\{-1,1\}^{\tilde N-2}}\Tr{i,j}{\hat H_{ij}^{}\Tr{[N]\setminus\{i,j\}} {\hat\Pi^{(\bm{k}_{i,j})}_{\bm{\alpha}\vphantom{l}}\,\ketbra{\psi_1}{\psi_2}\,\hat\Pi^{(\bm{k}_{i,j})}_{\bm{\alpha}\vphantom{l}}}}[\Big]\Big|
\nonumber\displaybreak[1]\\
&\leq\sum_{(i,j)\in{\cal I}_N^{(2)}} \sum_{\bm{\alpha}\in\{-1,1\}^{\tilde N-2}}
\Big|\Tr{i,j}{\hat H_{ij}^{}\Tr{[N]\setminus\{i,j\}}{\hat\Pi^{(\bm{k}_{i,j})}_{\bm{\alpha}\vphantom{l}}\ketbra{\psi_1}{\psi_2}
\,\hat\Pi^{(\bm{k}_{i,j})}_{\bm{\alpha}\vphantom{l}}}}[\Big]\Big|\nonumber\displaybreak[1]\\
&\leq\sum_{(i,j)\in{\cal I}_N^{(2)}} \sum_{\bm{\alpha}\in\{-1,1\}^{\tilde N-2}}
\|\hat H_{ij}^{}\|\cdot\big\|\Tr{[N]\setminus\{i,j\}}
{\hat\Pi^{(\bm{k}_{i,j})}_{\bm{\alpha}\vphantom{l}}\ketbra{\psi_1}{\psi_2}
\,\hat\Pi^{(\bm{k}_{i,j})}_{\bm{\alpha}\vphantom{l}}}\big\|_1\nonumber\displaybreak[1]\\
&\leq\max_{(i,j)\in{\cal I}_N^{(2)}}\|\hat H_{ij}^{}\| \sum_{(i,j)\in{\cal I}_N^{(2)}} \sum_{\bm{\alpha}\in\{-1,1\}^{\tilde N-2}}
\big\|\Tr{[N]\setminus\{i,j\}}{\hat\Pi^{(\bm{k}_{i,j})}_{\bm{\alpha}\vphantom{l}}\ketbra{\psi_1}{\psi_2}
\,\hat\Pi^{(\bm{k}_{i,j})}_{\bm{\alpha}\vphantom{l}}}\big\|_1\nonumber\displaybreak[1]\\
&\leq\max_{(i,j)\in{\cal I}_N^{(2)}}\|\hat H_{ij}^{}\| \sum_{(i,j)\in{\cal I}_N^{(2)}} \sum_{\bm{\alpha}\in\{-1,1\}^{\tilde N-2}}
\big\|\hat\Pi^{(\bm{k}_{i,j})}_{\bm{\alpha}\vphantom{l}}\ketbra{\psi_1}{\psi_2}
\,\hat\Pi^{(\bm{k}_{i,j})}_{\bm{\alpha}\vphantom{l}}\big\|_1\nonumber\displaybreak[1]\\
&=\max_{(i,j)\in{\cal I}_N^{(2)}}\|\hat H_{ij}^{}\| \sum_{(i,j)\in{\cal I}_N^{(2)}} \sum_{\bm{\alpha}\in\{-1,1\}^{\tilde N-2}}
\big\|\hat\Pi^{(\bm{k}_{i,j})}_{\bm{\alpha}\vphantom{l}}\ket{\psi_{1\vphantom{l}}^{}}\big\|\cdot
\big\|\hat\Pi^{(\bm{k}_{i,j})}_{\bm{\alpha}\vphantom{l}}\ket{\psi_{2\vphantom{l}}^{}}\big\|.
\end{align}\vspace{-4mm}
\end{widetext}
In the second inequality we used H\"older's inequality for the operator spectral and 1-norm $\big|\Tr{}{\hat X\hat Y}\big|\leq\|\hat X\|\cdot\|\hat Y\|_1$. In the third inequality we utilized the property $\|\Tr{1}{\hat X_{12}}\|_1\leq\|\hat X_{12}\|_1$ for $\hat X_{12}\in{\cal L}({\cal H}_1\otimes{\cal H}_2)\vphantom{\sum_0^{1/2}}$ \cite{SM_Watrous}, and finally in the last one we used $\big\|\ketbra{u}{v}\big\|_1=\|\ket{u}\|\cdot\|\ket{v}\|$. Now, from the factorization of joint probabilities we get
\begin{align}
&\big\|\hat\Pi^{(\bm{k}_{i,j})}_{\bm{\alpha}\vphantom{l}}\ket{\psi_l^{}}\big\|=\big\|\hat\Pi^{(k_1)}_{\alpha_1\vphantom{l}}\hat \Pi^{(k_2)}_{\alpha_2\vphantom{l}}\ldots\hat\Pi^{(k_{\tilde N-2})}_{\alpha_{\tilde N-2}\vphantom{l}}\ket{\psi_l^{}}\big\|\nonumber\\
&=(p_{l,\alpha_1}\,p_{l,\alpha_2}\ldots p_{l,\alpha_{\tilde N-2}})^{1/2}=(p_{l,+})^\frac{m}{2}(p_{l,-})^\frac{\tilde N-2-m}{2},
\end{align}
where $m$ is the number of positive eigenvalues among $\alpha_1,\alpha_2,\ldots,\alpha_{\tilde N-2}$. Hence, we find
\begin{widetext}\vspace{-3mm}
\begin{align}
|\bra{\psi_2}\hat H\ket{\psi_1}|&\leq\max_{(i,j)\in{\cal I}_N^{(2)}}\|\hat H_{ij}^{}\| \sum_{(i,j)\in{\cal I}_N^{(2)}} \sum_{\bm{\alpha}\in\{-1,1\}^{\tilde N-2}}
(p_{1,+}\,p_{2,+})^\frac{m}{2}(p_{1,-}\,p_{2,-})^\frac{\tilde N-2-m}{2}\nonumber\\
&\leq |\mbox{\;\!}{\cal I}_N^{(2)}|\max_{(i,j)\in{\cal I}_N^{(2)}}\|\hat H_{ij}^{}\|\sum_{\bm{\alpha}\in\{-1,1\}^{\tilde N-2}}
(p_{1,+}\,p_{2,+})^\frac{m}{2}(p_{1,-}\,p_{2,-})^\frac{\tilde N-2-m}{2}\nonumber\\
&=|\mbox{\;\!}{\cal I}_N^{(2)}|\max_{(i,j)\in{\cal I}_N^{(2)}}\|\hat H_{ij}^{}\|\;\sum_{m=0}^{\tilde N-2}\binom{\tilde N-2}{m}
(p_{1,+}\,p_{2,+})^\frac{m}{2}(p_{1,-}\,p_{2,-})^\frac{\tilde N-2-m}{2}\nonumber\\
&=|\mbox{\;\!}{\cal I}_N^{(2)}|\max_{(i,j)\in{\cal I}_N^{(2)}}\|\hat H_{ij}^{}\|\: q^{\tilde N-2},
\end{align}\vspace{-2mm}
\end{widetext}
where $q=\sqrt{p_{1,+}\,p_{2,+}}+\sqrt{p_{1,-}\,p_{2,-}}$. Since by assumption $\avg{\hat A^{(k)}}_{\psi_2}<\avg{\hat A^{(k)}}_{\psi_1}$, we have $p_{1,\pm}\neq p_{2,\pm}$. Using the inequality $\sqrt{xy}<(x+y)/2$ for distinct positive numbers, we obtain
\begin{align}
q&<\frac{p_{1,+}+p_{2,+}}{2}+\frac{p_{1,-}+p_{2,-}}{2}=1.
\end{align}
In addition, due to $\tilde N=O(N)$ for a macroscopic superposition state, we find the estimate $|\bra{\psi_2}\hat H\ket{\psi_1}| \leq\exp[-O(N)]$. Thus, we have derived an exponential bound with respect to $N$. In the general case of $K$-local Hamiltonian one could analogously derive the bound
\begin{align}
|\bra{\psi_2}\hat H\ket{\psi_1}|&\leq |\mbox{\;\!}{\cal I}_N^{(K)}|\max_{(i_1,\ldots,i_K)\in{\cal I}_N^{(K)}}\|\hat H_{i_1\ldots i_K}^{}\|\: q^{\tilde N-K}.
\end{align}
Based on this result, the Lemma implies that the energy gap as well vanishes exponentially fast in the macroscopic limit $N\to\infty$.

\subsection{Proof of the Theorem}

Our task is to estimate the magnitude of the matrix element $\langle\psi_2|\hat H|\psi_1\rangle$ of a 2-local Hamiltonian, under the assumptions from the main text. Recall that an additive observable $\hat S=\sum_{i=1}^N\hat S_i$ introduces the decomposition of a given state $\ket{\psi}=\sum_{m=1}^M\sqrt{p_m}\ket{s_m}$, where $\hat S\ket{s_m}=s_m\ket{s_m}$, $\sqrt{p_m}\ket{s_m}=\sum_{|\bm{\sigma}| =s_m}\sum_{\bm{\mu}}\ket{\bm{\sigma},\bm{\mu}}\braket{\bm{\sigma},\bm{\mu}}{\psi}$ and $|\bm{\sigma}|\equiv\sum_{i=1}^N\sigma_i$. Furthermore, we used the separating eigenvalue $s_{\bar m}$ to express the ground state in the form of superposition $\ket{\psi}=a_1\ket{\psi_1}+a_2\ket{\psi_2}$, with $a_1\ket{\psi_1}=\sum_{m=1}^{\bar{m}-1}\sqrt{p_m}\ket{s_m}$ and $a_2\ket{\psi_2}=\sum_{m=\bar{m}}^M\sqrt{p_m}\ket{s_m}$. Then, we find the following
\begin{widetext}\vspace{-3mm}
\begin{align}\label{eq:H12:p}
\bra{\psi_2}&\hat H\ket{\psi_1}=\sum_{(i,j)\in{\cal I}_N^{(2)}} \bra{\psi_2}\hat H_{ij}^{}\ket{\psi_1}=\frac{1}{a_1 a_2}\sum_{(i,j)
\in{\cal I}_N^{(2)}}\sum_{m=1}^{\bar{m}-1}\sum_{m'=\bar{m}}^M\sqrt{p_{m'}}\:\!\bra{s_{m'}}\hat H_{ij}^{}\ket{s_m}\sqrt{p_m}.
\end{align}
\end{widetext}
Evaluation of $\bra{s_{m'}}\hat H_{ij}^{}\ket{s_m}$ boils down to considering $\bra{\bm{\sigma}',\bm{\mu}'}\hat H_{ij}^{}\ket{\bm{\sigma},\bm{\mu}}=\Tr{i,j}{\hat H_{ij}^{}\Tr{[N]\setminus\{i,j\}}{\ketbra{\bm{\sigma}, \bm{\mu}}{\bm{\sigma}',\bm{\mu}'}}}$, where $|\bm{\sigma}'|=s_{m'}$ and $|\bm{\sigma}|=s_m$. One finds that
\begin{widetext}\vspace{-2mm}
\begin{align}
\Tr{[N]\setminus\{i,j\}}{\ketbra{\bm{\sigma}, \bm{\mu}}{\bm{\sigma}',\bm{\mu}'}}=\ket{\sigma_i^{},\mu_i^{}}^{}_i\ket{\sigma_j^{},\mu_j^{}}^{}_j\bra{\sigma_i',\mu_i'}^{}_i\bra{\sigma_j',\mu_j'}^{}_j\prod_{k\in [N]\setminus\{i,j\}}\delta_{\sigma_k',\sigma_k^{}}\delta_{\mu_k',\mu_k^{}}.
\end{align}\vspace{-2mm}
\end{widetext}
A necessary condition for the last product to be nonzero is $\sum_{k\in [N]\setminus\{i,j\}}(\sigma_k'-\sigma_k^{})=0 \vphantom{\frac{1}{\sum_\otimes}}$, i.e., $s_{m'}-s_m=\sigma_i'+\sigma_j'-(\sigma_i^{}+\sigma_j^{})$. Since $-2\delta_\varsigma\leq \sigma_i'+\sigma_j'-(\sigma_i^{}+\sigma_j^{})\leq 2\delta_\varsigma$, all the nonvanishing terms from (\ref{eq:H12:p}) must obey $-2\delta_\varsigma\le s_{m'}-s_m\le 2\delta_\varsigma$, while by construction we have $s_{m'}\geq s_{\bar m}$ and $s_m<s_{\bar m}$. Hence, the only nonzero terms are those related to the triangular region ${\cal T}_{\bar{m}}$ in the $(m,m')$-plane that is determined by the previous inequalities. Let $m_>$ be the largest $m$ such that $s_m < s_{\bar m}+2\delta_\varsigma$. Similarly, let $m_<$ be the smallest $m$ such that $s_m\geq s_{\bar m}-2\delta_\varsigma$. We obtain the following
\begin{widetext}\vspace{-4mm}
\begin{subequations}\label{eq:|H12|}
\begin{align}
|\bra{\psi_2}\hat H\ket{\psi_1}|&=\frac{1}{a_1 a_2}\,\Big|\!\sum_{(i,j)\in{\cal I}_N^{(2)}}\sum_{(m,m')\in{\cal T}_{\bar m}}\sqrt{p_{m'}} \:\!\bra{s_{m'}}\hat H_{ij}^{}\ket{s_m}\sqrt{p_m}\,\Big|\nonumber\displaybreak[1]\\
&=\frac{1}{a_1 a_2}\,\Big|\!\sum_{(i,j)\in{\cal I}_N^{(2)}}\sum_{m=m_<}^{\bar{m}-1}\sum_{m'=\bar{m}}^{m_>}\sqrt{p_{m'}}\:\!\bra{s_{m'}}\hat H_{ij}^{}\ket{s_m}\sqrt{p_m}\,\Big|\displaybreak[1]\\
&=\frac{1}{a_1 a_2}\,\Big|\!\sum_{(i,j)\in{\cal I}_N^{(2)}}\sum_{m=m_<}^{\bar{m}-1}\sum_{m'=\bar{m}}^{m_>}\:\Tr{i,j}{\hat H_{ij}^{}\Tr{[N]\setminus\{i,j\}}{\sqrt{p_m}\ketbra{s_m}{s_{m'}}\sqrt{p_{m'}}}}[\Big]\Big|\nonumber\displaybreak[1]\\
&=\frac{1}{a_1 a_2}\,\Big|\!\sum_{(i,j)\in{\cal I}_N^{(2)}}\:\Tr{i,j}{\hat H_{ij}^{}\Tr{[N]\setminus\{i,j\}} {\sum_{m=m_<}^{\bar{m}-1}\sqrt{p_m}\ket{s_m}\sum_{m'=\bar{m}}^{m_>}\sqrt{p_{m'}}\bra{s_{m'}}}}[\Big]\Big|\nonumber\displaybreak[1]\\
&\le\frac{1}{a_1 a_2}\sum_{(i,j)\in{\cal I}_N^{(2)}}\Big|\:\!\Tr{i,j}{\hat H_{ij}^{}\Tr{[N]\setminus\{i,j\}} {\sum_{m=m_<}^{\bar{m}-1}\sqrt{p_m}\ket{s_m}\sum_{m'=\bar{m}}^{m_>}\sqrt{p_{m'}}\bra{s_{m'}}}}[\Big]\Big|\nonumber\displaybreak[1]\\
&\le\frac{1}{a_1 a_2}\sum_{(i,j)\in{\cal I}_N^{(2)}}\|\hat H_{ij}^{}\|\cdot\Big\|\:\!\Tr{[N]\setminus\{i,j\}} {\sum_{m=m_<}^{\bar{m}-1}\sqrt{p_m}\ket{s_m}\sum_{m'=\bar{m}}^{m_>}\sqrt{p_{m'}}\bra{s_{m'}}}\Big\|_1\displaybreak[1]\\
&\le\frac{1}{a_1 a_2}\sum_{(i,j)\in{\cal I}_N^{(2)}}\|\hat H_{ij}^{}\|\cdot\Big\|\sum_{m=m_<}^{\bar{m}-1} \sqrt{p_m}\ket{s_m}\sum_{m'=\bar{m}}^{m_>}\sqrt{p_{m'}}\bra{s_{m'}}\Big\|_1\displaybreak[1]\\
&\le\frac{|\mbox{\;\!}{\cal I}_N^{(2)}|}{a_1 a_2}\max_{(i,j)\in{\cal I}_N^{(2)}}\|\hat H_{ij}^{}\|\cdot\Big\|\sum_{m=m_<}^{\bar{m}-1} \sqrt{p_m}\ket{s_m}\sum_{m'=\bar{m}}^{m_>}\sqrt{p_{m'}}\bra{s_{m'}}\Big\|_1\nonumber\displaybreak[1]\\
&=\frac{|\mbox{\;\!}{\cal I}_N^{(2)}|}{a_1 a_2}\max_{(i,j)\in{\cal I}_N^{(2)}}\|\hat H_{ij}^{}\|\cdot\Big\|\sum_{m=m_<}^{\bar{m}-1} \sqrt{p_m}\ket{s_m}\Big\|\cdot\Big\|\sum_{m'=\bar{m}}^{m_>}\sqrt{p_{m'}}\bra{s_{m'}}\Big\|\displaybreak[1]\\
&=\frac{|\mbox{\;\!}{\cal I}_N^{(2)}|}{a_1 a_2}\max_{(i,j)\in{\cal I}_N^{(2)}}\|\hat H_{ij}^{}\|\cdot\Big(\sum_{m=m_<}^{\bar{m}-1} p_m\Big)^{1/2}\cdot\Big(\sum_{m'=\bar{m}}^{m_>}p_{m'}\Big)^{1/2}\nonumber\displaybreak[1]\\
&\le\frac{|\mbox{\;\!}{\cal I}_N^{(2)}|}{2\:\!a_1 a_2}\max_{(i,j)\in{\cal I}_N^{(2)}}\|\hat H_{ij}^{}\|\cdot\Big(\sum_{m=m_<}^{\bar{m}-1} p_m+\sum_{m'=\bar{m}}^{m_>}p_{m'}\Big)\displaybreak[1]\\
&\le\frac{|\mbox{\;\!}{\cal I}_N^{(2)}|}{2\:\!a_1 a_2}\max_{(i,j)\in{\cal I}_N^{(2)}}\|\hat H_{ij}^{}\|\cdot P_\psi(|s-s_{\bar m}|\le 2\delta_\varsigma).
\end{align}
\end{subequations}\vspace{-4mm}
\end{widetext}
In the line (\ref{eq:|H12|}a) we found convenient to extend the summation over ${\cal T}_{\bar m}$ to the summation over the encompassing rectangular region. Note that all the added terms are actually zero-terms. Thereafter, $m$ and $m'$ index the eigenvalues of $\hat S$ within the interval $[s_{\bar m}- 2\delta_\varsigma,s_{\bar m})$ and $[s_{\bar m},s_{\bar m}+2\delta_\varsigma)$, respectively. The line (\ref{eq:|H12|}b) is a consequence of H\"older's inequality for operator spectral and \mbox{1-norm} $\big|\Tr{}{\hat X\hat Y}\big|\leq\|\hat X\|\cdot\|\hat Y\|_1$, whereas the line (\ref{eq:|H12|}c) follows from $\|\Tr{1}{\hat X_{12}}\|_1\leq\|\hat X_{12}\|_1$ for $\hat X_{12}\in{\cal L}({\cal H}_1\otimes{\cal H}_2)$ \cite{SM_Watrous} and we used $\|\mbox{\;\!}\ketbra{u}{v}\mbox{\:\!}\|_1=\|\ket u\|\cdot\|\ket v\|$ in the line (\ref{eq:|H12|}d). Finally, in the line (\ref{eq:|H12|}e) we invoked the inequality $\sqrt{x y}\leq(x+y)/2$ for nonnegative reals. Recalling the Lemma from the main text, the fact $\braket{\psi_2}{\psi_1}=0$ and the choice $E_0=0$, we find
\begin{widetext}\vspace{-4mm}
\begin{align}\label{eq:Gap}
\Delta E\le\frac{|\mbox{\;\!}{\cal I}_N^{(2)}|}{2\:\!a_1^2 a_2^2}\max_{(i,j)\in{\cal I}_N^{(2)}}\|\hat H_{ij}^{}\|\cdot P_\psi(|s-s_{\bar m}|\le 2\delta_\varsigma).
\end{align}
\end{widetext}
The proof of the Theorem is now completed.

\subsection{Example of $W$ state $\ket{j,j-1}$}

\begin{figure}[b]
\centering
\includegraphics[width=\linewidth]{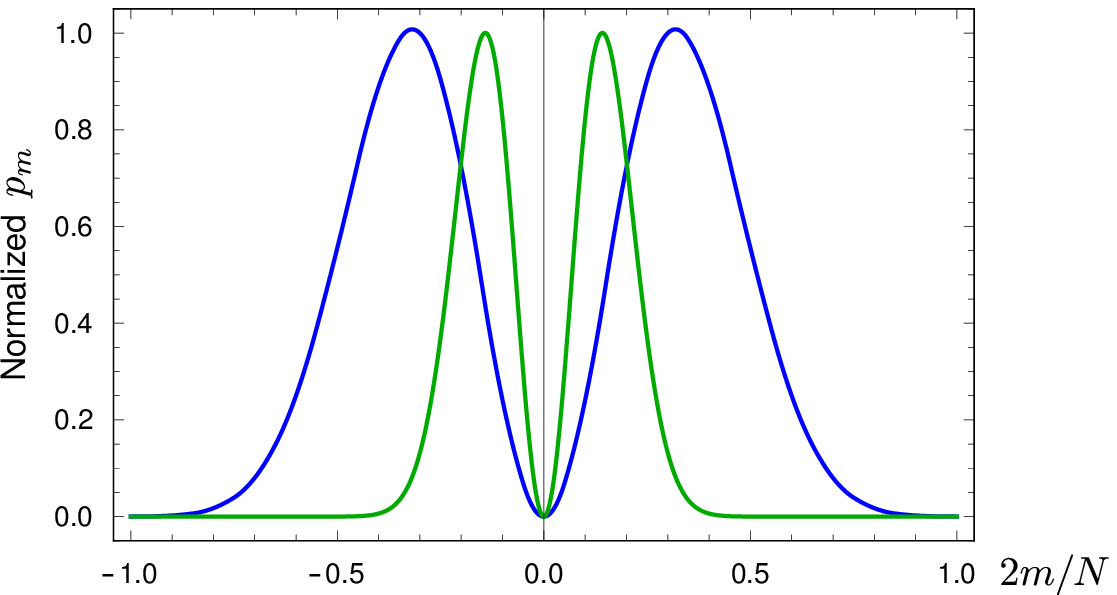}
\caption{Normalized probability distributions $p_m$ for $W$ state for $N=40$ (blue line) and $N=200$
(green line). Normalization is such that the maximal value is unity. Continuous curves are used for aesthetic purposes.}
\label{fig:2}
\end{figure}

Here, we give the derivation of the probability distribution $p_m=|\braket{j,j-1}{j,m}_x|^2$ for $j\to\infty$ and small $m$, where $\ket{j,m}_x=e^{-i\frac{\pi}{2}\hat J_y}\ket{j,m}$. Particular example of such distribution is given in Fig.\ \ref{fig:2}. First, we will evaluate the overlap
\begin{align}
\braket{j,j-1}{j,m}_x&=\frac{1}{\sqrt{2j}}\bra{j,j}\hat J^+\ket{j,m}_x\nonumber\\
&=\frac{1}{\sqrt{2j}}\bra{j,j}(\hat J_x+i\hat J_y)e^{-i\frac{\pi}{2}\hat J_y}\ket{j,m}\nonumber\\
&=\frac{1}{\sqrt{2j}}\bra{j,j}e^{-i\frac{\pi}{2}\hat J_y}(\hat J_z+i\hat J_y)\ket{j,m},
\end{align}
where we used properties and definition of the angular momentum ladder operator $\hat J^+$, as well as the relation $e^{i\frac{\pi}{2}\hat J_y}\hat J_x\:\!e^{-i\frac{\pi}{2}\hat J_y}=\hat J_z$. Next, we employ
\begin{eqnarray}
0&=&\bra{j,j}\hat J^-e^{-i\frac{\pi}{2}\hat J_y}\nonumber\\
&=&\bra{j,j}(\hat J_x-i\hat J_y)e^{-i\frac{\pi}{2}\hat J_y}\nonumber\\
&=&\bra{j,j}e^{-i\frac{\pi}{2}\hat J_y}(\hat J_z-i\hat J_y),
\end{eqnarray}
so that
\begin{align}\label{eq:Jz_iJy}
\bra{j,j}e^{-i\frac{\pi}{2}\hat J_y}\hat J_z=\bra{j,j}e^{-i\frac{\pi}{2}\hat J_y}\:\!i\hat J_y,
\end{align}
and we conclude
\begin{align}\label{eq:j,j-1}
\braket{j,j-1}{j,m}_x&=\frac{1}{\sqrt{2j}}\bra{j,j}e^{-i\frac{\pi}{2}\hat J_y}\:\!2\hat J_z\ket{j,m}\nonumber\\
&=m\sqrt{\frac{2}{j}}\:\!\bra{j,j}e^{-i\frac{\pi}{2}\hat J_y}\ket{j,m}\equiv m\sqrt{\frac{2}{j}}\,c_m.
\end{align}
In order to calculate the matrix element, denoted by $c_m$, we proceed as follows. First, from (\ref{eq:Jz_iJy}) and the relation $2i\hat J_y=\hat J^+ -\hat J^-$, we get
\begin{align}\label{eq:cs1}
2m\:\!c_m&=\sqrt{(j\!-\!m)(j\!+\!m\!+\!1)}\,c_{m+1}\nonumber\\
&-\sqrt{(j\!+\!m)(j\!-\!m\!+\!1)}\,c_{m-1}.
\end{align}
Second, using $e^{i\frac{\pi}{2}\hat J_y}\hat J_z\mbox{\:\!}e^{-i\frac{\pi}{2}\hat J_y}=-\hat J_x$ together with $\hat J_x=(\hat J^+ +\hat J^-)/2$, we find
\begin{align}
\bra{j,j}\hat J_z\:\!e^{-i\frac{\pi}{2}\hat J_y}&=-\bra{j,j}e^{-i\frac{\pi}{2}\hat J_y}\hat J_x\nonumber\\
&=-\frac{1}{2}\bra{j,j}e^{-i\frac{\pi}{2}\hat J_y}(\hat J^+ +\hat J^-),
\end{align}
which allows us to obtain
\begin{align}\label{eq:cs2}
2j\:\!c_m&=-\sqrt{(j\!-\!m)(j\!+\!m\!+\!1)}\,c_{m+1}\nonumber\\
&-\sqrt{(j\!+\!m)(j\!-\!m\!+\!1)}\,c_{m-1}.
\end{align}
From the two relations (\ref{eq:cs1}) and (\ref{eq:cs2}), we derive the recurrence relation $c_m=-\:\!\sqrt\frac{j+m+1}{j-m}\,c_{m+1}$,
which leads to $c_m=(-1)^{j-m}\sqrt{\binom{2j}{j+m}}\,c_j$. Using the normalization condition $\sum_{m=-j}^j|c_m|^2 = 1$, we get
$|c_j|=\frac{1}{2^j}$ and $|c_m|=\frac{1}{2^j}\sqrt{\binom{2j}{j+m}}$. Finally, from (\ref{eq:j,j-1}) we obtain
\begin{align}
p_m=\frac{2m^2}{2^{2j}\:\!j}\binom{2j}{j+m},
\end{align}
as stated in the main text. The asymptotic behavior for fixed $m$ and $j\to\infty$ can be easily obtained using Stirling's asymptotic series.

\subsection{Example of superpositions of Dicke states}

Additionally, we will demonstrate that certain superpositions of Dicke states cannot be unique ground states of 2-local Hamiltonians. We consider $N=2j$ spin-$1/2$ particles, with $j$ integer (for the notational simplicity). Thus, $d=2$, $\ell=2$, and $\delta_\varsigma=1$. Assume that the unique ground state of some 2-local Hamiltonian of the spins has the following form
\begin{align}\label{eq:psi_n}
\ket{\psi_n^\pm}=\sum_{k=0}^n (\pm)^k c_k\ket{j,-n+2k},\quad n=O(N^0)\in\mathbb{N},
\end{align}
where the coefficients $c_k\in\mathbb{C}$ satisfy $\sum_{k=0}^n |c_k|^2=1$ and $\sum_{k=0}^n c_k=0$. Some particular instances of such
states are $(\ket{j,-1}\mp\ket{j,1})/\sqrt{2}$, $(\ket{j,-2}\mp\ket{j,4})/\sqrt{2}$, $(\ket{j,-3}-2\ket{j,1}+\ket{j,5})/\sqrt{6}$, etc.
It can be verified that the proper additive observable for $\ket{\psi_n^+}$ states is $\hat J_y$, while for $\ket{\psi_n^-}$ states it is
$\hat J_x$. All the states (\ref{eq:psi_n}) are in fact general macroscopic quantum states since the variance of the additive observable
scales as $O(N^2)$.

\begin{figure}[t]
\centering
\includegraphics[width=0.95\linewidth]{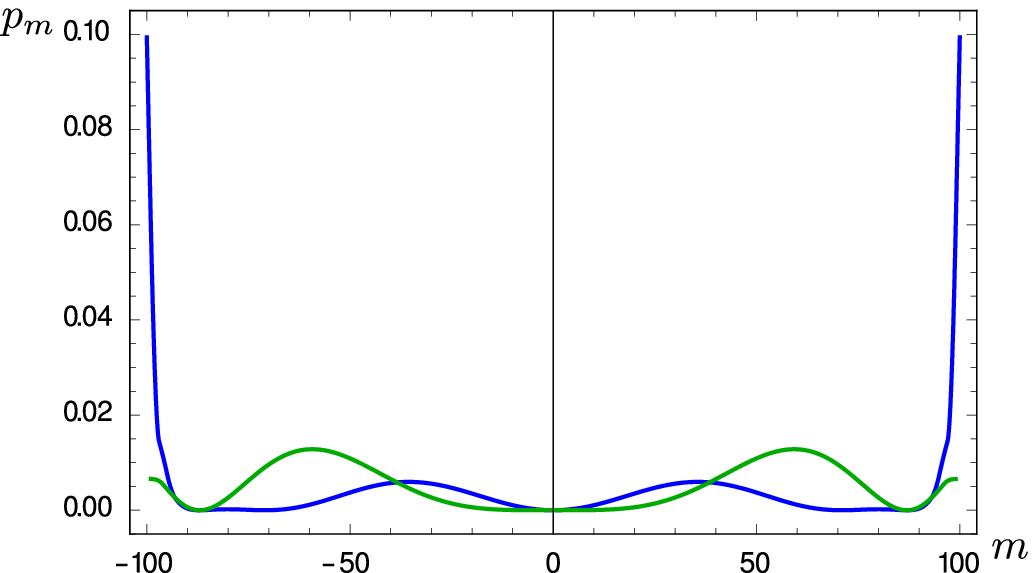}
\caption{The probability distribution $p_m$ for the superposition $(\ket{j,-1}+\ket{j,5})/\sqrt{2}$ of two Dicke states for
$j=100$. The blue (green) line labels $p_m$ for $m$ even (odd). Continuous curves are used for aesthetic purposes.}
\label{fig:3}
\end{figure}

We first concentrate on $\ket{\psi_n^-}$ states. The required probability distribution is given by
$p_m=|\braket{\psi_n^-}{j,m}_x|^2$, $s_m=m$, and we again select $s_{\bar m}=0$ (see Fig.\ \ref{fig:3} for an example).
We are going to analyze the behavior of the probabilities $p_m$ for $j\to\infty$ and small $m$, i.e., we want to examine the overlap
$\braket{j,m'}{j,m}_x=\bra{j,m'}e^{-i\frac{\pi}{2}\hat J_y}\ket{j,m}=d^j_{m'm}(\pi/2)\vphantom{\frac{1}{\sum}}$ for $j$ large. In the last
equality we recognized the Wigner (small) $d$ function that can be related to Jacobi polynomials $P^{(a,b)}_n(z)$ in the
following manner \cite{SM_AMQP}
\begin{align}
d^j_{m'm}(\theta)&=\left[\frac{(j+m)!(j-m)!}{(j+m')!(j-m')!}\right]^\frac{1}{2} P^{(m-m',m+m')}_{j-m}(\cos\theta)\nonumber\\
&\times \left(\sin\frac{\theta}{2}\right)^{\!\!m-m'}\left(\cos\frac{\theta}{2}\right)^{\!\!m+m'}.
\end{align}
Thus, we find
\begin{align}
d^j_{m'm}\left(\frac{\pi}{2}\right)\!=\!\frac{1}{2^m}\!\left[\frac{(j+m)!(j-m)!}{(j+m')!(j-m')!}\right]^\frac{1}{2}
\!\!P^{(m-m',m+m')}_{j-m}(0).
\end{align}
Using Stirling's asymptotic series and asymptotic expansion of Jacobi polynomials \cite{SM_Frenzen85,SM_Wong04,SM_Bai07} in the limit $j\to\infty$
and $m,m'$ finite, we obtain
\begin{align}
\braket{j,m'}{j,m}_x\sim\sqrt\frac{2}{\pi j}\cos\frac{(j-m+m')\pi}{2}+O\big(\:\!j^{-3/2}\big).
\end{align}
so that
\begin{align}
&\braket{\psi_n^-}{j,m}_x=\sum_{k=0}^n(-1)^k c_k^*\braket{j,-n+2k}{j,m}_x\nonumber\\
&\sim\sqrt\frac{2}{\pi j}\sum_{k=0}^n(-1)^k c_k^*\cos\left[\frac{(j-m-n)\pi}{2}+k\pi\right]+O\big(\:\!j^{-3/2}\big)\nonumber\\
&=\sqrt\frac{2}{\pi j}\cos\frac{(j-m-n)\pi}{2}\sum_{k=0}^n c_k^*+O\big(\:\!j^{-3/2}\big)\nonumber\\
&=O\big(\:\!j^{-3/2}\big),
\end{align}
since by construction we have $\sum_{k=0}^n c_k=0\vphantom{\frac{1}{\sum}}$. Thus, we establish the asymptotic relation
$p_m=|\braket{\psi_n^-}{j,m}_x|^2=O(j^{-3})$. The choice $s_{\bar m}=0$ guaranties that $a_k\to 1/\sqrt{2}$ ($k=1,2$) as $j\to\infty$, so that the separation probability $P_\psi(|s-s_{\bar m}|\leq 2)$ vanishes at least as $O(j^{-3})$, i.e., $O(N^{-3})$. Essentially the same approach can also be applied to $\ket{\psi_n^+}$ states. Finally, we conclude that none of the states (\ref{eq:psi_n}) can be reached by cooling the system of $N$ spin-$1/2$ particles described by an arbitrary $2$-local Hamiltonian.

This example can also be put into the context of double-well (or twofold fragmented single-well) Bose-Einstein condensates of $N$ particles via the Schwinger representation of angular momentum operators in terms of two bosonic modes. Hence, for arbitrary pairwise particle interactions and potential trap designs the considered superpositions of Dicke states cannot in principle arise as a result of the simple condensation process and necessitate other means of preparation. Contrarily, we conclude from our previous analysis that one would need genuine $3$-particle interactions in order for it to be possible to prepare such states by the process of cooling.

\subsection{Quantum superpositions in SQUIDs}

Here we provide the discussion of superpositions of magnetic-flux states in SQUIDs and show the relation to our analysis. We will consider the simplest form of Josephson device which displays all the features relevant for the present discussion, namely a single rf SQUID \cite{SM_Leggett02}. In the thermodynamic limit (the number of Cooper pairs $N$ tends to infinity), the full many-body description reduces to a simple model with one macroscopic quantum variable, i.e., the total flux $\Phi$ trapped through the SQUID ring, and the dynamics follows an effective single-particle 1D Schr\"odinger equation, where the effective Hamiltonian $\hat H_{\rm eff}(\Phi)$ has a usual kinetic $\propto-\partial^2/\partial\Phi^2$ and a potential term $U(\Phi)$ \cite{SM_Eckern84}. The system exhibits a finite energy gap $\Delta E$ independent of $N$. For an appropriate choice of external magnetic field, the problem boils down to the analysis of a 1D quantum particle in a double-well potential $U(\Phi)$ \cite{SM_Leggett02}. The ground state wave function $\psi_0(\Phi)$ has two peaks to which we can associate the states $\psi_-(\Phi)$ and $\psi_+(\Phi)$. They correspond to the states of supercurrent flowing in one or in the other direction around the ring. Since the magnitude of the total magnetic moment in each of the cases can be $10^6\mu_B$ \cite{SM_VanDerWal00}, or even $10^{10}\mu_B$ \cite{SM_Friedman00}, these states are asserted to be macroscopically distinct.

For simplicity reasons, let us assume the symmetric potential $U(-\Phi)=U(\Phi)$ with two degenerate wells separated by a classically impenetrable barrier \cite{SM_Leggett02}. For the case of an even potential, the well-known textbook result states that the ground state wave function is even, i.e.\ $\psi_0(-\Phi)=\psi_0(\Phi)$, whereas the first excited state wave function is odd $\psi_1(-\Phi)=-\psi_1(\Phi)$. Here $\hat H_{\rm eff}(\Phi)\psi_i(\Phi)=E_i\psi_i(\Phi)$ and $\Delta E=E_1-E_0$. Since $\psi'_0(0)=0$, the probability density $|\psi_0(\Phi)|^2$ attains a minimum at the center of the barrier $\Phi=0$. Precisely this is the natural choice for the separation point that divides the ground state wave function into the two components $\psi_{\pm}(\Phi)$.

Following Ref.\ \cite{SM_Robnik00}, simple algebraic manipulation of eigenequations $\hat H_{\rm eff}(\Phi)\psi_i(\Phi)=E_i\psi_i(\Phi)$ yields the relation
\begin{equation}\label{gap}
\Delta E={\rm const}\times\frac{\psi_0(0)\psi_1'(0)}{\int_0^{\infty}\psi_0(\Phi)\psi_1(\Phi)d\Phi},
\end{equation}
meaning that the energy gap is directly proportional to the ground state probability amplitude $\psi_0(0)$ at the center of the barrier. Since $\Delta E$ is nonzero, $\psi_0(0)$ must be nonvanishing as well. Therefore, as long as the energy gap is finite, there is a nonvanishing macroscopic probability density $|\psi_0(0)|^2$ of Cooper pairs at the center of the barrier (the separation point). Thus, one concludes that the states $\psi_{\pm}(\Phi)$ cannot be arbitrarily well separated whenever the energy gap is finite. In addition, the same general conclusion as above holds for arbitrary confining potential. Namely, it is a well-known fact that a non-degenerate ground state wave function has no nodes, i.e., it exhibits the nonzero probability density everywhere.

Finally, we point out that instead of an effective description and an analysis of the flux variable, one might consider the full 2-local many-body Hamiltonian and invoke the analysis of some  additive observable, such as the pseudo-angular-momentum \cite{SM_DiRienzo82}. In such a case, the dependence on the number of Cooper pairs $N$ would explicitly be taken into account. Our main theorem would then directly yield the conclusion that for any considered additive observable there is a lower bound on the separation probability $P_\psi \ge O(1/N^2)$. In other words, this is the best separation of the two wave function components one can expect to have.

\end{document}